\documentclass[onecolumn]{jpsj3}
\usepackage{bm}
\usepackage{epsfig}
\usepackage{graphicx}
\usepackage{epstopdf}
\usepackage{color}
\usepackage{ulem}
\usepackage{cancel}
\def\urs{URu$_2$Si$_2$}

\def\runauthor{F. Bourdarot \textit{et al.}}

\begin{document}

\title{Precise study of the resonance at Q$_0$=(1,0,0) in \urs}

\author{Frederic Bourdarot\thanks{E-mail address: frederic.bourdarot@cea.fr}, Elena Hassinger, Stephane Raymond, Dai Aoki, Valentin Taufour, Louis-Pierre Regnault, Jacques Flouquet}
\inst{INAC, SPSMS, CEA-Grenoble, 38054 Grenoble, France}

\date{\today}

\abst{New inelastic neutron scattering experiments have been performed on \urs\ with special focus on the response at \textbf{Q$_0$}=(1,0,0), which is a clear signature of the hidden order (HO) phase of the compound. With polarized inelastic neutron experiments, it is clearly shown that below the HO temperature ($T_0=17.8$ K) a collective excitation (the magnetic resonance at $E_0\simeq1.7$ meV) as well as a magnetic continuum co-exist. Careful measurements of the temperature dependence of the resonance lead to the observation that its position shifts abruptly in temperature with an activation law governed by the partial gap opening and that its integrated intensity has a BCS-type temperature dependence. Discussion with respect to recent theoretical development is made.}

\kword{\urs, Hidden Order, magnetic resonance, PrRu$_4$P$_{12}$, PrFe$_4$P$_{12}$}

\maketitle

\section{Introduction}

The puzzling nature of the hidden order (HO) phase of \urs\ is still not understood. The central interest of this enigma is that it reflects the duality between the local and itinerant characters of the $5f$ electrons. These different facets often play a major role in the field of strongly correlated electronic systems. In spite of more than two decades of intense search\cite{Palstra:1985}, there is still no direct access to the order parameter (OP) as it occurs for the sublattice magnetization of the high pressure antiferromagnetic (AF) ground state with the wave-vector \textbf{Q$_{AF}$}=(0,0,1)\cite{Amitsuka:1999,Bourdarot:2005} which appears above $P_x\simeq0.5$ GPa via a first order transition switching from HO to AF phases\cite{Motoyama:2003,Amitsuka:2007,Hassinger:2008}. However, recently it was noticed that an unambiguous signature of the HO phase is the sharp resonance at $E_0\simeq1.8$ meV for the commensurate wave-vector \textbf{Q$_0$}=(1,0,0) (equivalent wave-vector to \textbf{Q$_{AF}$}) as this resonance mode collapses through $P_x$ while the other resonance at $E_1\simeq4.1$ meV for the incommensurate wave-vector \textbf{Q$_1$}=(1.4,0,0) persists through $P_x$\cite{villaume:2008}. Furthermore, above $P_x$, a magnetic field leads to the ''resurrection'' of the resonance at $E_0$, when the HO phase is restored \cite{aoki:2009}. As the strong inelastic signal at \textbf{Q$_0$} is replaced above $P_x$ by a large elastic signal, fingerprint of the AF ground state with \textbf{Q$_{AF}$}=(0,0,1), it was proposed that, in both HO and AF phases, a lattice doubling along the \textbf{c} axis occurs at the transition from paramagnetic (PM) to either HO or AF ground states\cite{aoki:2009}. The nice feedback is that the change in the class of tetragonal symmetry via development of a new Brillouin zone generates a drastic decrease in the carrier number as pointed out by a large number of theories\cite{Yamagami:2000,Harima:2009,Elgazzar:2009} and experiments\cite{Hassinger:2008a}. Furthermore, as the HO-AF line touches the PM-HO and PM-AF lines at a critical pressure $P_c\simeq 1.4$ GPa and critical temperature $T_c\simeq19.5$ K, a supplementary symmetry breaking must occur between the HO and AF phases. Two recent theoretical proposals for the OP of the HO phase were a hexadecapole - from Dynamical Mean Field Theory (DMFT) calculations\cite{Haule:2009};  or a $O_{xy}$ type antiferroquadrupole - from group theory analysis \cite{Harima::2010}. For both models, the pressure-switch from HO to AF will add a supplementary time reversal breaking in the AF phase. For these models as well as for recent band structure calculations \cite{Elgazzar:2009}, partial gapping of the Fermi surface may happen at $T_0=17.8$ K with a characteristic gap $\Delta_G$. It was even proposed in the last model, that in the HO phase the gapping is produced by a spontaneous symmetry breaking occurring through collective AF moment excitations.

This article presents a careful revisit of the inelastic neutron response at \textbf{Q$_0$} \cite{Broholm:1991,Mason:1995,Wiebe:2007} using recent progress in polarized inelastic neutron configuration of the spectrometer IN22 and in the performance of the cold-neutron three-axis spectrometer IN12, both installed at the high flux reactor of the Institute Laue Langevin (ILL). Thus, this new generation of experiments provide a careful basis on the temperature and pressure evolution of $E_0$. They give a new insight on previous neutron studies thanks to our recent proof\cite{villaume:2008} that the resonance at $E_0$ is up to now the major signature of the HO phase. Comparison will be made with the case of charge ordering observed in the skutterudite PrRu$_4$P$_{12}$ as well as with PrFe$_4$P$_{12}$ where a sequence of HO-AF phases are observed \cite{JPSJ77SA:2008}.

\section{Experimental set up}

High quality single crystals of \urs\ were grown by the Czochralski method in tetra-arc furnace. The details are described elsewhere \cite{aoki:2010}. The sample already used for inelastic neutron scattering in the superconducting state \cite{Hassinger:2010}, was installed in an ILL-type orange cryostat with $\bf{c}$ axis oriented vertically on the thermal triple-axis IN22 in its polarized configuration and on the cold triple-axis IN12 in its standard configuration (both CEA-CRG spectrometer at ILL).

The IN22 experiment was performed at 1.5 K, with two fixed final energies : 14.7 meV ($k_f$=2.662 \AA$^{-1}$) and 30.3 meV ($k_f$=3.84 \AA$^{-1}$). The beam was polarized by a Heusler monochromator vertically focusing and analyzed in energy and polarization by a Heusler analyzer vertically (fixed) and horizontally focusing. The flipping ratio was around 17 and the energy resolutions were 0.95 meV and 2.4 meV for both energies, respectively. No collimation was installed. The background was optimized by an optical calculation of the dimension openings of the slits placed before and after the sample. The measurements were performed in the non-spin-flip channel which provides all the necessary information. The inelastic scans were performed with $\bf{Q}$ parallel to the $\bf{a}$ axis. When the polarization is along the $\bf{a}$ axis, the intensity measured (I$^a_{NSF}$) corresponds to the nuclear and background contributions; when the polarization is along the $\bf{b}$ or $\bf{c}$ axis, the intensities measured (I$^b_{NSF}$) and (I$^c_{NSF}$) correspond to the same contributions plus the (imaginary part of the) susceptibility along the $\bf{b}$ or $\bf{c}$ directions. In the following, the data shown are the subtraction of inelastic scans performed in the non-spin-flip channel with polarization along the $\bf{b}$ or $\bf{c}$ axis, with an inelastic scan performed in the same conditions with polarization along the $\bf{a}$ axis. The subtractions correspond to the imaginary part of the dynamical susceptibilities with for the $\bf{b}$ axis $\chi''_y\propto (I^b_{NSF}-I^a_{NSF})$ the transverse susceptibility and for the $\bf{c}$ axis $\chi''_z\propto (I^c_{NSF}-I^a_{NSF})$ the longitudinal susceptibility.

The  IN12 experiment was performed with a fixed final energy $E_f=4.7$ meV ($k_f$=1.5 \AA$^{-1}$) that gives a good compromise between intensity of the excitation and energy resolution $l=0.11$ meV. The incident neutrons were selected by a (0,0,2) graphite vertically focusing monochromator with vertically focusing and analyzed in energy by a (0,0,2) graphite analyzer with horizontal focusing. No collimation was installed. The temperature was measured by a calibrated carbon thermometer and it was checked before each scan that the temperature was stable. As for IN22, the background was optimized by an optical calculation of the dimension openings of the slits placed before and after the sample. The raw scans were corrected for the electronic background (29 counts per hour) and for the $\lambda/2$ contamination of the monitor.

\section{Description of the model used for fitting the inelastic spectrum}
In a neutron scattering experiment, the neutron intensity $I(q,\omega)$ in the detector is proportional to the convolution of the scattering function $S(q,\omega)$ with the instrumental resolution function. $S(q,\omega)$  is related to the imaginary part of the dynamical spin susceptibility $\chi''(q,\omega)$ ($\chi(q,\omega)= \chi'(q,\omega)+\imath\chi''(q,\omega)$) via the fluctuation-dissipation theorem : $S(q,\omega)=n(\omega,T)\chi''(q,\omega)$  where $n(\omega,T)= 1/(1-e^{-\hbar\omega/{k_BT}})$ is the detailed balance factor. 
 Below $T_{0}$, the excitation is well-defined with an asymmetrical shape related to the finite extent of the resolution function that integrates the dispersion of the excitation in the vicinity of the nominal $q$ wave-vector at which the spectrometer is set-up. To analyze the data, firstly a harmonic oscillator function is taken for $\chi"(q,\omega$) ; this corresponds to the difference of two normalized Lorentzian functions multiplied by $\frac{\chi_o\,\Omega^2_0}{\omega_0\gamma}$, where $\chi_0=\chi(q,\omega=0)$ is the static susceptibility, $\Omega_0$ is the oscillator frequency, $\gamma$ its damping, and $\omega_0$ is given by the equation $\Omega_0=\sqrt{\omega^2_0+(\gamma/2)^2}$. Secondly, a simplified convolution with the resolution function is made : to this aim the resolution function is approximated by a 4D parallelepiped (instead of an ellipsoid) and the $q$ dispersion is taken as linear in all directions. This linear dispersion simplified the calculation and gives a better description of the dispersion than a usual quadratic law in the case of \urs. This description is called the \bm{$\gamma$} \textbf{model} when $\gamma \gg l$ and the \bm{$l$} \textbf{model} when $l \gg \gamma$, $l$ being the energy resolution. It leads to a simple analytic expression for $I(q,\omega)$ at the minimums of the dispersion ($q_0$):

\begin{equation}
I(q_0,\omega)=n(\omega,T)\left(F(q_0,\omega,)-F(q_0,-\omega)\right)\\
\label{m3}
\end{equation}
with
\begin{eqnarray}
\begin{split}
F(q_0,\omega)&=L\,K(q_0,\omega)\left(\sum_{i=x,y,z}\sqrt{1+(2\alpha_i\,l/\omega_0(q_0))^2}\right.\nonumber\\
&\left.e^{-\left(4\ln{2}\left(\alpha_i\frac{\omega-\omega_0(q_0)}{\omega_0(q_0)}\right)^2\right)} \right)\\
K(q_0,\omega)&=\frac{1}{2}\left(1+\frac{2}{\pi}arctan\left(\frac{2\,(\omega-\omega_0(q_0))}{\beta}\right)\right)\nonumber\\
\end{split}
\end{eqnarray}
\\

\noindent
where $\beta=\gamma(q_0)$ for the \bm{$\gamma$} \textbf{model} and $\beta=l$ for the \bm{$l$} \textbf{model} 
$\alpha_i=\alpha_{i0}(\omega_0/\Omega_0)^2$, where $\alpha_{i0}$ is the ratio between the slope of the dispersion and the $q$-width of the resolution function. The L parameter, which depends of the magnetic form factor, the incident  and final energies ($k_i$ and $k_f$) is supposed to stay constant at first approximation.

Finally the susceptibility $\chi_{0}$ for $T<T_0$ is determined using the integrated intensity by applying the formula:

\begin{equation}
\begin{split}
I_{\Omega_0}=\int^\infty_0\chi''(q_0,\omega)d\omega=\\
\frac{\chi_0 \Omega^2_0(q_0)}{\omega_0(q_0)}\arctan\left(\frac{\omega_0(q_0)}{\gamma(q_0)/2}\right)
\end{split}
\label{m4}
\end{equation}

Above $T_{0}$, the signal is much broader than the resolution and no convolution is needed (at least on IN12). For $\chi"(q,\omega)$, we use the magnetic quasi-elastic model, which corresponds to a Lorentzian function of susceptibility $\chi_{L0}$ and the full-width at half-maximum $\Gamma_L$, multiplied by $\omega$:

\begin{equation}
\chi''(q_0,\omega)=\frac{\chi_{L0}\,\omega\,\Gamma_L}{\omega^2+\Gamma^2_L}
\label{lor}
\end{equation}

 Between \bm{$\gamma$} \textbf{model} and the quasi-elastic model, the widths and the susceptibilities are linked by the relations: $\Gamma_L=\gamma/2$ and $\chi_0=\chi_{L0}$.

\section{Results}

\subsection{Inelastic Polarized neutron scattering}

At first this experiment was to unambiguously determine the polarization of the resonances at \textbf{Q$_0$} = (1,0,0) and \textbf{Q$_1$} = (1.4,0,0) and to determine the nature of the signal occurring at much higher energy than the resonant modes. The origin of such a signal is often referred to as coming from multi-phonons. However, it may come from magnetic process as was hinted by C. Broholm \cite{Broholm:1991}.

\begin{figure}[htb]
\begin{center}
\includegraphics[width=80mm]{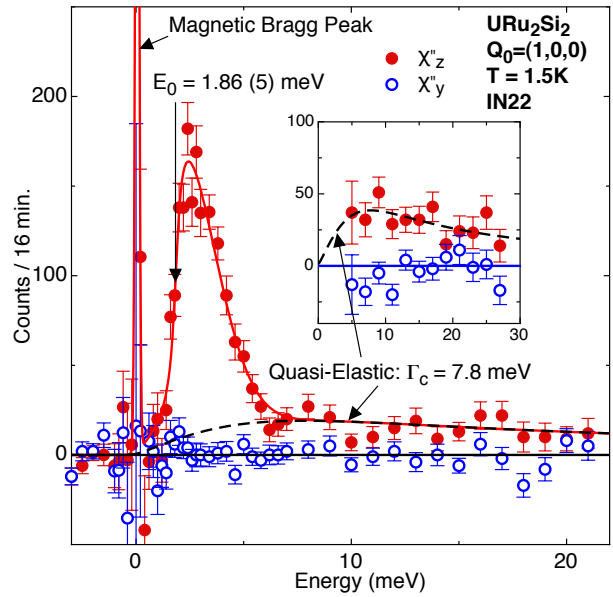}
\end{center}
\caption{(color online) Transverse (open circles) and longitudinal (filled circles) magnetic response of \urs\ at \textbf{Q$_0$} and $T=1.5$ K with a final energy $E_f=14.7$ meV.  The full and dashed curves correspond to a \bm{$l$}\textbf{-model} of eq.(\ref{m3}) and quasi-elastic function for the magnetic continuum, respectively. The vertical black arrow indicates the gap position of the resonance (1.86(5) meV) for the \bm{$l$}\textbf{-model}. The inset shows the same magnetic response but with  a final energy $E_f=30.3$ meV.}
\label{fig5}
\end{figure}

Figure \ref{fig5} shows the longitudinal and transverse magnetic response at \textbf{Q$_0$} measured on IN22 with a final energy $E_f=14.7$ meV. No transverse magnetic response is detected for a range of energy transfer going from -3 meV to  27 meV. The longitudinal magnetic response shows two contributions: the well-known and well-defined resonance with a gap value $E_0$ fitted to 1.86(5) meV, where $E_0$ is the harmonic oscillator energy $\Omega_{0}$  (the gap value $E_0$ measured with the cold triple-axis spectrometer IN12 is around 1.70(5) meV) and a broad magnetic contribution. This broad magnetic contribution, which looks like to a magnetic continuum persists at least up to 27 meV as seen in the inset of Figure \ref{fig5} for measurements performed with a final energy of 30.3 meV (27\ meV was the maximum energy transfer we could reach in this configuration). In this paper the intensity of this continuum is described by a Lorentzian function (eq.\ref{lor}) of width $\Gamma_{c}$. An extra elastic signal (at $\omega=0$) corresponding to the small antiferromagnetic moment is detected. This well known signal is currently believed to be a parasitic contribution due to the survival of AF droplets generated near defects \cite{Matsuda:2001,Amato:2004}. 

\begin{figure}
\begin{center}
\includegraphics[width=80mm]{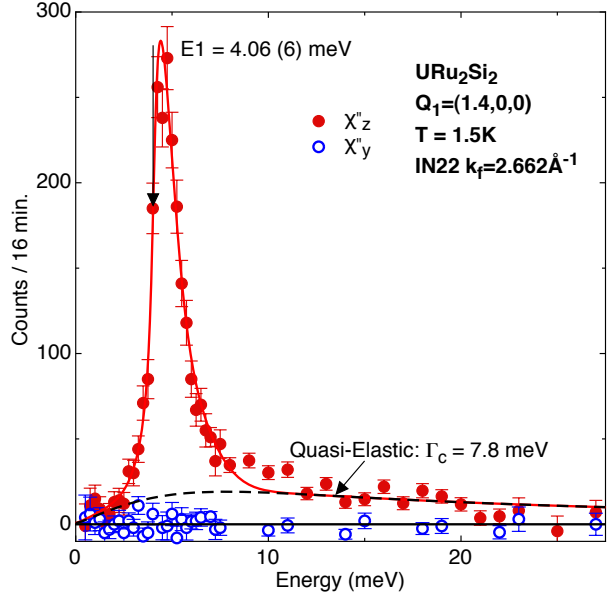}
\end{center}
\caption{Transverse (open circles) and longitudinal (filled circles) response of \urs\ at \textbf{Q$_1$} and $T=1.5$ K. The full and dashed curves correspond to a \bm{$l$}\textbf{-model} of eq.(\ref{m3}) and quasi-elastic function for the magnetic continuum, respectively. The vertical black arrow indicates the gap position of the resonance (4.06(6) meV) for the \bm{$l$}\textbf{-model}.}
\label{fig6}
\end{figure}

Figure \ref{fig6} shows the longitudinal and transverse magnetic response at \textbf{Q$_1$} measured with a final energy $E_f=14.7$ meV.  As previously at \textbf{Q$_0$}, no transverse magnetic response is detected in the range of 0 meV to 27 meV. Again the longitudinal response shows two inelastic magnetic contributions: the well-known and well-defined resonance with a gap $E_1$ of 4.06(6) meV, and the broad continuum of linewidth $\Gamma_c$. Note that the gap $E_1=4.06(6)$ meV is slightly lower than the usual gap value $E_1\simeq4.5$ meV found in previous experiment \cite{Bourdarot:2003}. Let us point out that it is the first time that the resolution and the dispersion are taken into account to analyse this excitation. 

\begin{figure}
\begin{center}
\includegraphics[width=80mm]{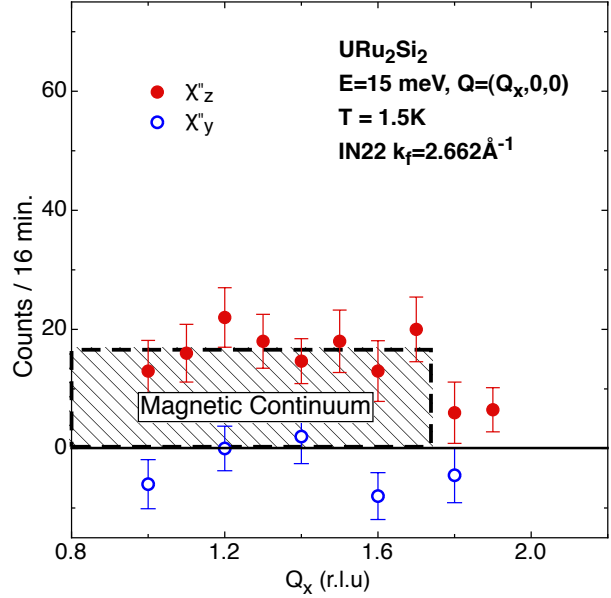}
\end{center}
\caption{Transverse (open circles) and longitudinal (filled circles) response of \urs\ at $T=1.5$ K for a \textbf{Q}-scan (Q$_x$,0,0) at 15 meV. The hatched area corresponds to the magnetic continuum.}
\label{fig8}
\end{figure}

Figure \ref{fig8} shows the longitudinal and transverse magnetic response for a \textbf{Q}-scan performed with an energy transfer of 15 meV. A constant signal corresponding to the continuum is measured from \textbf{Q}=(1,0,0) to \textbf{Q}=(1.7,0,0), then the magnetic signal decreases approaching to nuclear zone center \textbf{Q}=(2,0,0). The vanishing of the continuum has to be verified by new \textbf{Q}-scans at different energy transfer in the futur.

To summarize, the inelastic polarized neutron scattering experiments performed at \textbf{Q$_0$} and \textbf{Q$_1$} confirm without any ambiguity that, at low temperature (below $T_0=17.8$ K), the magnetic response is exclusively longitudinal. We also evidence a broad magnetic continuum that may be fitted at least for the two main \textbf{Q} positions (\textbf{Q$_0$} and \textbf{Q$_1$}) by exactly the same quasi-elastic function with a half-width $\Gamma_c=7.8$ meV. This magnetic contribution was never taken into account in the previous studies of temperature dependence of the gaps at \textbf{Q$_0$} and at \textbf{Q$_1$}. This motivates us to reinvestigate the temperature dependence of the excitation at \textbf{Q$_0$}.

\subsection{Low energy study at \textbf{Q$_0$}}

The aim of this study is a precise determination of the temperature dependence of the magnetic resonance at the position \textbf{Q$_0$}. This experiment was already performed by T.E. Mason \cite{Mason:1995b} but without a large precision and more important without taking into account the magnetic continuum described by a quasi-elastic function and revealed by our recent inelastic polarized neutron scattering.  From their study they concluded that the gap follows a singlet ground state model but moreover that the evolution of the susceptibility $\chi(\mathbf{Q}_0,\omega=0)$ shows a large enhancement at $T_0$. We will discuss these points below.

\begin{figure}
\begin{center}
\includegraphics[width=80mm]{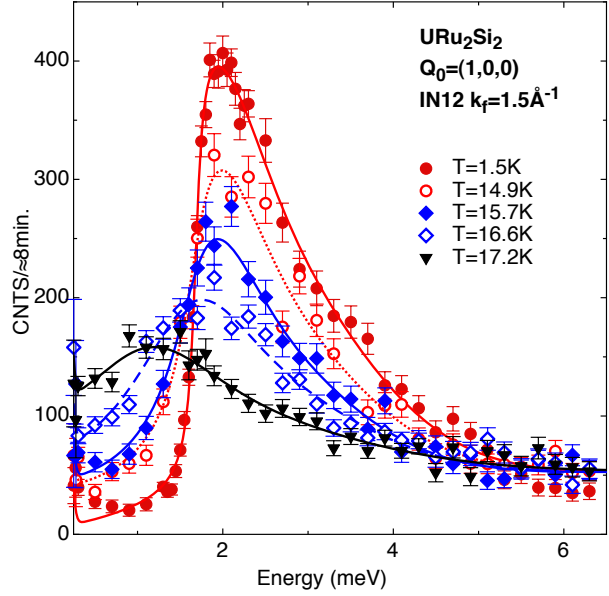}
\end{center}
\caption{Energy scans measured at \textbf{Q$_0$} for temperatures below $T_0=17.8$ K. Fits are performed with the \bm{$\gamma$} \textbf{model} plus the continuum (quasi-elastic with $\Gamma_c=7.8$ meV).}
\label{fig11}
\end{figure}

\begin{figure}
\begin{center}
\includegraphics[width=80mm]{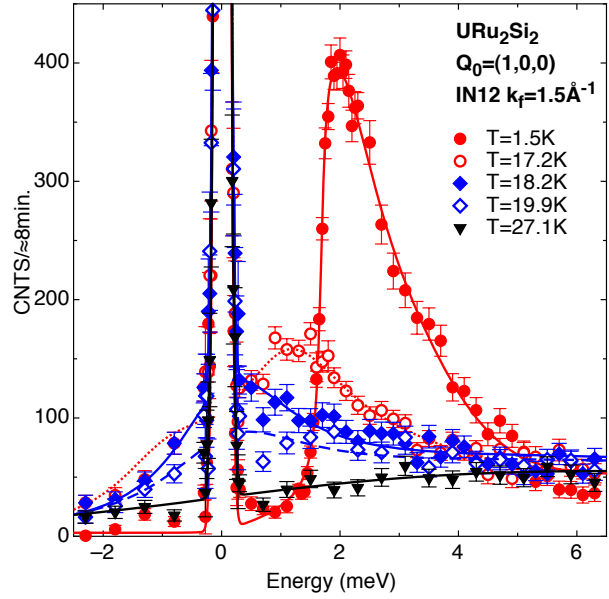}
\end{center}
\caption{Energy scans measured at \textbf{Q$_0$} for temperatures just around the transition temperature $T_0=17.8$ K (and at 1.5 K for reference). Fits are described in the text. At $T=27. 1$ K (triangles), the scan is only fit by the continuum (quasi-elastic signal with $\Gamma_c=7.8$ meV).}
\label{fig12}
\end{figure}

Figures \ref{fig11} and \ref{fig12} show some representative inelastic spectrums obtained at \textbf{Q$_0$} just below and above the transition temperature $T_0=17.8$ K respectively. At $T$=1.5 K, the signal measured at energies above 5-6 meV is derived from the continuum previously detected by inelastic polarized neutron scattering. As the half-width $\Gamma_c=7.8$ meV of the quasi-elastic function which fit the continuum is already known at low temperature, and assuming that for temperatures lower than $\Gamma_c/k_B \approx 88$ K, this width does not change, only the amplitude can depend on temperature. A good fit is obtained by taking a constant amplitude for all temperatures: the continuum being, at $T=27.1$ K, the unique contribution (as seen in Fig. \ref{fig12}).

\begin{figure}
\begin{center}
\includegraphics[width=80mm]{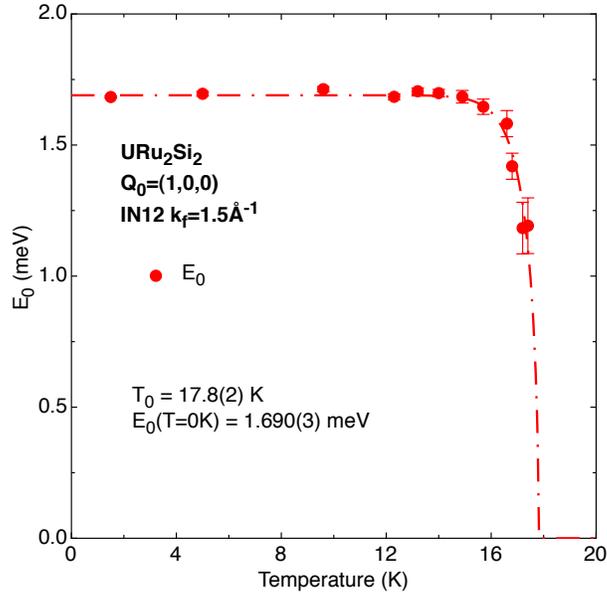}
\end{center}
\caption{Temperature dependence of the gap $E_0$. The curves are guide for the eyes.}
\label{fig10}
\end{figure}

\begin{figure}
\begin{center}
\includegraphics[width=80mm]{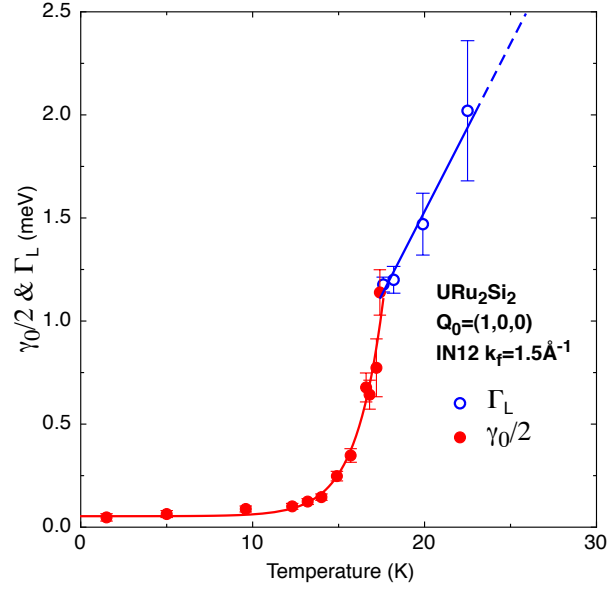}
\end{center}
\caption{Temperature dependence of the half-width $\gamma_0$ of the resonance at \textbf{$Q_0$} below $T_0$ (filled circle) and of the quasi-elastic $\Gamma_L$ above $T_0$ (open circle). The curve below $T_0$ corresponds to a fit with a Korringa model.}
\label{fig14}
\end{figure}

For $T<T_0$ (Fig. \ref{fig11}), in parallel to the continuum, the well-defined resonance is detected as for $T=1.5$ K. The inelastic spectrums show clearly that the width $\gamma_0$ and the gap $E_0$ change substantially with temperature only close to $T_0$. The gap $E_0$ and the width $\gamma_0$ determined using the \bm{$\gamma$} \textbf{model} are shown in Fig. \ref{fig10} and \ref{fig14}. As expected from the spectrum, the gap $E_0$ decreases only close to $T_0$ but more abruptly as the width $\gamma_0$ increases. At $T_0$, $E_0\approx\gamma_0/2$, that confirms that we enter into an over-damped regime for temperatures larger than $T_0$.

For $T>T_0$ (Fig. \ref{fig12}), the magnetic response is over-damped, and the spectrum is treated with a quasi-elastic magnetic function of width $\Gamma_L$ added to the continuum ($\Gamma_c=7.8$ meV). As expected, $\Gamma_L=\gamma_0/2$ at $T_0$, which validates the \bm{$\gamma$} \textbf{model} and indicates that the gap $E_0$ drops to zero in the paramagnetic state (above $T_0$).  The half-width of this excitation is plotted versus temperature in Figure \ref{fig14} (open circles). $\Gamma_L$ increases rapidly (maybe linearly) with temperature. It was not possible to follow this signal for temperatures approaching $T=27.5$ K: above this temperature only the large tail of the continuum can be detected.

\begin{figure}
\begin{center}
\includegraphics[width=80mm]{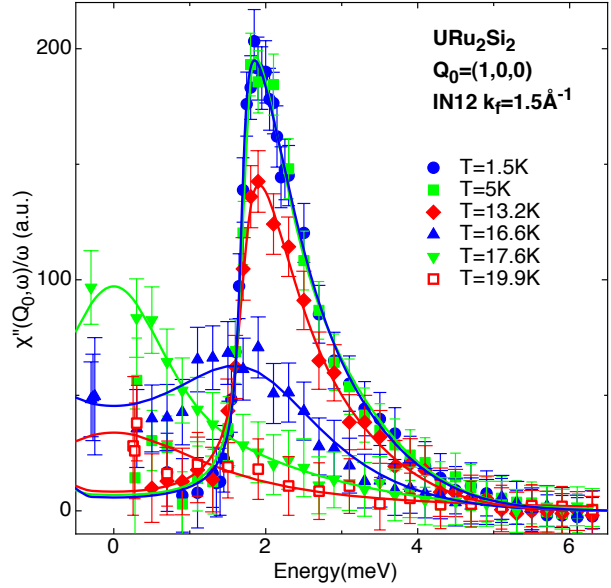}
\end{center}
\caption{Imaginary part of the dynamical spin susceptibility without the magnetic continuum divided by energy $(\chi''(\mathbf{Q}_0,\omega)/\omega$) for different temperatures. The curves are deduced from the $\gamma$ model corrected by the thermal factor and divided by $\omega$. The integration of these curves gives $\chi(\textbf{Q$_0$},\omega=0)$.}
\label{fig20}
\end{figure}

As the temperature evolution of the signal is not usual, the Figure \ref{fig20} gives the variation of $\chi''(\mathbf{Q}_0,\omega)/\omega$ as a function of $\omega$ at different temperatures. Thus, this plot shows that this signal saturates at low temperature and the abrupt drop of E$_0$ when approaching $T_0$. Clearly the integration of $\chi''(\mathbf{Q}_0,\omega)/\omega$ increases on cooling below $T_0$ as discussed latter as a consequence of Fermi Surface reconstruction.

\begin{figure}
\begin{center}
\includegraphics[width=80mm]{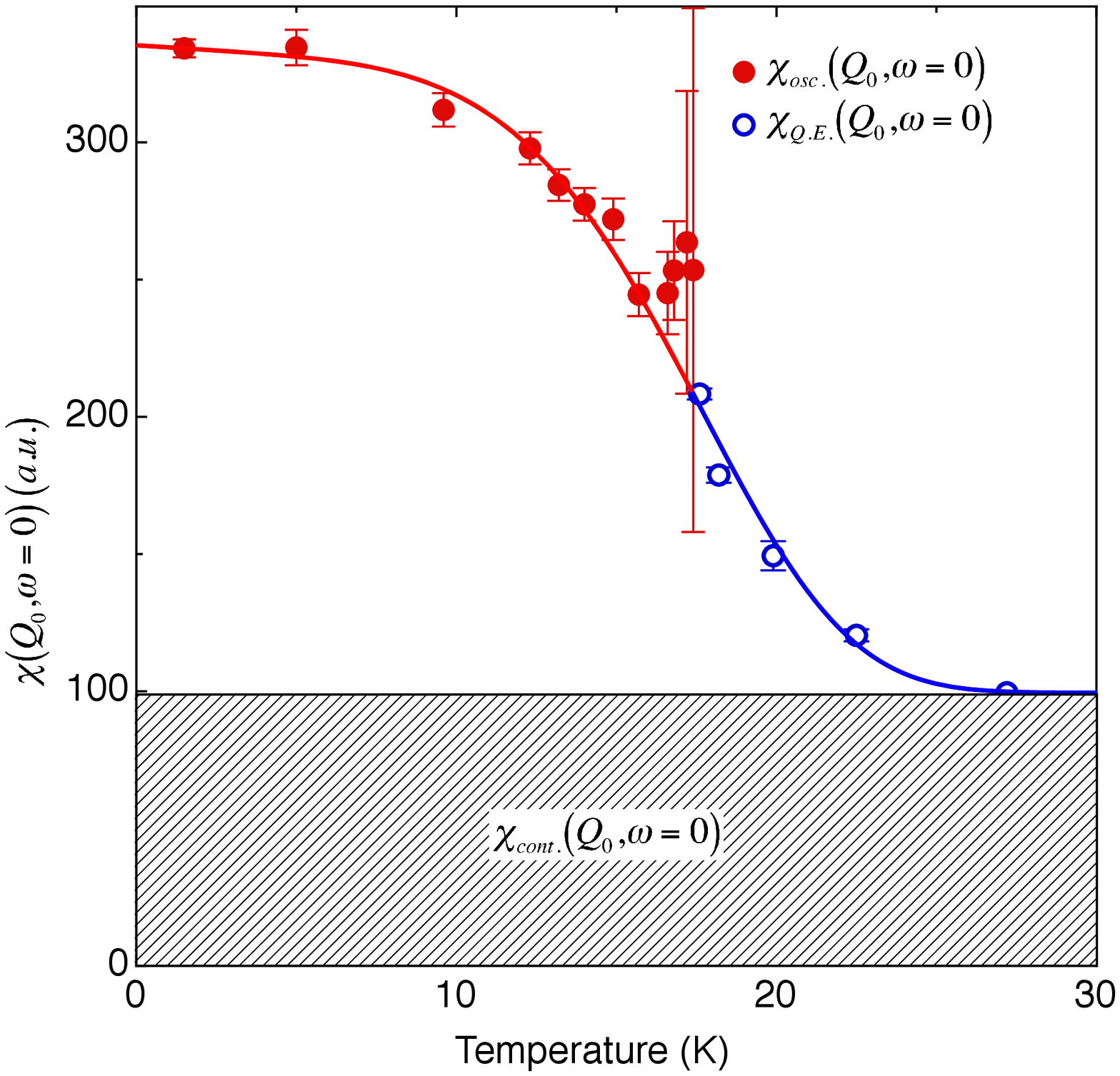}
\end{center}
\caption{Temperature dependence of the static susceptibility $\chi(\textbf{Q}_0,\omega=0)$. In filled circles are data coming from the \bm{$\gamma$} \textbf{model}, the open circles from the quasi-elastic model $\Gamma_L$, and the hatched area coming from the magnetic continuum ($\Gamma_c=7.8$ meV).}
\label{fig15}
\end{figure}

Figure \ref{fig15} shows the magnetic susceptibility at $Q_0$, $\chi(\mathbf{Q}_0,\omega=0)$. The susceptibility for $T<T_0$ is determined using eq. (\ref{m4}) (filled circles), then $\chi_{L0}$ is a fitted parameter of the quasi-elastic expression for $T>T_0$ (open circles). Of course, in addition, the susceptibility coming from the magnetic continuum (with $\Gamma_c=7.8$ meV) has to be added (constant low contribution in Fig \ref{fig15}). Without any scaling-factor, the susceptibilities from the \bm{$\gamma$} \textbf{model} and the second magnetic quasi-elastic contribution ($\Gamma_L$) are equal at $T_0$. The total susceptibility versus temperature shows a saturation at low temperature, then decreases  but with a bump around $T_0$. The susceptibility stays almost constant above $T = 30$ K. In contrast to Mason's analysis no marked divergence of $\chi(\mathbf{Q}_0,\omega=0)$ is observed at $T_0$. Furthermore, in our data below 16 K, we found that $\chi(\mathbf{Q}_0,\omega=0)$ increases on cooling before saturating at low temperature.

\begin{figure}
\begin{center}
\includegraphics[width=80mm]{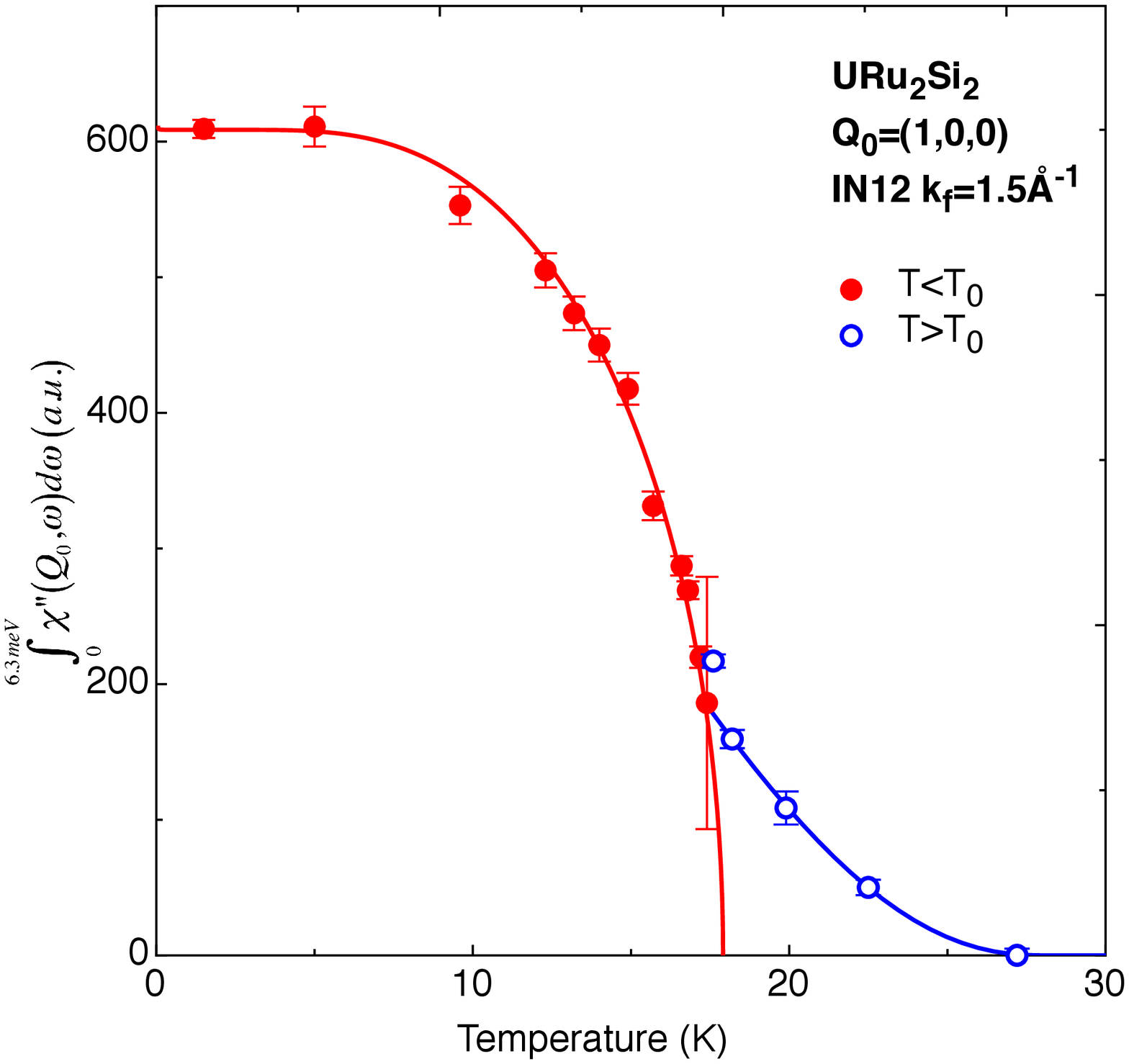}
\end{center}
\caption{Temperature dependence of the integrated imaginary part of the dynamical spin susceptibility. In filled circles are data coming from the \bm{$\gamma$} \textbf{model}, the open circles from the quasi-elastic model. The intensity coming from the magnetic continuum ($\Gamma_c=7.8$ meV) was not taken into account. The line is the BCS-type gap below $T_0$, and a guide for the eyes above $T_0$.}
\label{fig13}
\end{figure}

The integration $\int^{6.3 meV}_0\chi''(\mathbf{Q}_0,\omega)d\omega$ ($\simeq I_{E_0}$)  of the magnetic resonance at \textbf{$Q_0$} without the continuum contribution ($\Gamma_c$) is plotted in Fig. \ref{fig13}. The filled circles correspond to the contribution with the oscillator model and the open circles correspond to the integration of the quasi-elastic contribution of width $\Gamma_L$. This integration, below $T_0$, seems to mimic the temperature variation of an OP vanishing at $T_0$, while above $T_0$ it behaves like the contribution of a critical regime. Furthermore, the temperature variation of $I_{E_0}(T)$ is well described below $T_0$ by BCS formula used for the temperature variation of the superconducting gap \cite{Bardeen:1957}.

\section {Discussion}

The occurrence of a magnetic continuum ($\Gamma_c = 7.8$ meV) for energies $\omega$ far higher than the energy $k_BT_0$ of the hidden order phase is a general phenomenon: large energy transfer leads to excite energy states not only from the ground state.

As the study is focused on the hidden order phase, we have concentrated our studies on the frequency response at \textbf{Q$_0$}. Below $T_0$, the resonance at $E_0$ represents a collective mode with a half-width $\gamma_0/2$ much smaller than the gap energy $E_0$: the collective mode is long lifetime as shown by the large ratio $\frac{E_0}{\gamma_0/2}\approx35$. As shown on Fig. \ref{fig10}, in decreasing temperature $E_0$ very rapidly reaches its final value $\approx1.7$ meV. On crossing $T_0$, the resonance collapses and the magnetic response appears quasi-elastic with a half-width $\Gamma_L$ becoming rapidly larger than the resonance energy $E_0$ at $T = 0$ K.

The susceptibility $\chi($\textbf{Q$_0$},$\,\omega$=0) reported in Fig. \ref{fig15} is calculated for $T<T_0$ using eq. (\ref{m4}). Due to the weakness of the resonance signal close to $T_0$ the uncertainty in the determination of  $\chi($\textbf{Q$_0$},$\,\omega$=0) is large. However it is obvious that no divergence  of  $\chi($\textbf{Q$_0$},$\,\omega$=0) occurs at $T_0$ as it is observed at the onset of an antiferromagnetic ordering. In heavy fermion compounds close to AF-PM critical point, as in Ce$_{1-x}$La$_x$Ru$_2$Si$_2$ series, a sharp maximum of $\chi(\textbf{Q$_0$},\omega=0$) occurs at $T_N$ (for x$<$0.075) which is accompanied on cooling , below $T_N$ by a decrease of $\chi(\textbf{Q$_0$},\omega=0)$ \cite{Knafo:2009}. Here on cooling below $T_0$, by contrast, $\chi(\textbf{Q$_0$},\omega=0)$ increases at it is observed for critical concentration x$_c=0.075$ in Ce$_{0.925}$La$_{0.075}$Ru$_2$Si$_2$. However, the major difference is that for \urs\ the saturation regime is achieved very steeply in temperature as the key energy is the gap energy $\Delta_G$ (see Figure \ref{fig10}). Also a drastic difference between \urs\ and CeRu$_2$Si$_2$ cases is that the resonance at E$_0$ collapses suddenly above P$_x$ at the benefit of the establishment of a large sublattice magnetization. The occurrence of a weak maximum of $\chi(\textbf{Q$_0$},\omega=0)$ at $T_0$ may be consequence of the Fermi surface reconstruction with the particularity that the U ions through $T_0$ will go from an intermediate valence behavior above $T_0$ to a quasi-tetravalent dressing below $T_0$  \cite{Hassinger:2008a}. This image, based on thermodynamical considerations, is supported by the recent tunneling microscope experiments which appear during the revision of our paper \cite{aynajan:2010,schmidt:2010}. Thus this observation confirms the lack of antiferromagnetism at \textbf{Q$_0$} in the hidden order phase. Thanks to our previous measurements\cite{villaume:2008}, we have specified the emergence of the resonance at \textbf{Q$_0$}=(1,0,0) as the signature of the hidden order (see also \cite{Ohkawa:1999,Miyake:2010,Miyake:2010b}), we interpret the temperature dependence $I_{E_0}(T)$ plotted in Fig. \ref{fig13} as the temperature dependence of the OP\cite{villaume:2008}.
It was suggested by P.M. Oppeneer according the model developed in reference \cite{Elgazzar:2009,Oppeneer:2009} that $I_{E_0}(T)$ may be related with the hidden order parameter via an even function of the magnetization amplitude which will not vanish in time even in the hidden order phase as the sublattice magnetization amplitude for an AF.

As the quasi-elastic contribution with the characteristic energy half-width $\Gamma_L$ above $T_0$ seems to vanish above $T\approx30$ K, a simple picture would be that only the continuum with $\Gamma_c=7.8$ meV persists above $T=30$ K. According to the Kondo impurity model which predicts \cite{1}
$\Gamma\gamma=750$ mJ mol$^{-1}$K$^{-2}$, the linear term of the specific heat of $\gamma$ can be approximated to $\approx 96$ mJ mol$^{-1}$K$^{-2}$ with the continuum $\Gamma_c$. This agrees with the observed magnitude of the linear term of the specific heat \cite{vanDijk:1997b}. Of course an open problem is the modification of this continuum at low energy when crossing $T_0$ since transport measurements, as well as NMR and specific heat measurements, indicate clearly a Fermi surface reconstruction with a carrier drop by a factor 3 to 10 \cite{Schoenes:1987,Kasahara:2007,Behnia:2005,Bel:2004,Kohara:1986,Kohara:1987,Maple:1986}, which favors localized magnetism.

A key point is the rapid temperature evolution of the energy gap $E_0$ which is much faster than the temperature evolution of the BCS-type gap. The emerging image is that, at $T_0$, the change in lattice symmetry associated with the paramagnetic to hidden order phase transition (from $body-centered\ tetragonal$ to $simple\ tetragonal$) induced presumably by a multipolar ordering leads to large gapping of the Fermi surface with characteristic gap energy $\Delta_G$ much larger than $E_0$. The origin of $E_0$ may come from crystal-field splitting with dispersion coming from dipolar and quadrupolar interactions (see for example \cite{JPSJ77SA:2008,Kuramoto:2009,Balatsky:2009}). Opening a gap $\Delta_G$ at $T_0$ leads to a drop of the carrier number which allows the observation and development of the resonance: the resonance at \textbf{Q$_0$} is over-damped far above $T_0$ become a well-defined excitation below $T_0$. Of course the change of carrier number acts as well on the over-damped mode observed above $T_0$ for \textbf{Q}$_1$=(1.4,0,0), allowing again the appearance of a sharp resonance below $T_0$ at $E_1\approx4$ meV \cite{Broholm:1991,Wiebe:2007}. 
Since $E_1$ is quite comparable to $\Gamma_c$ below $T_0$, this explains why the inelastic contribution at \textbf{Q$_1$} is still observed in the over-damped regime, above $T_0$. It was suggested that the hidden order phase may be an incommensurate antiferromagnet at \textbf{Q}$_1$ \cite{Balatsky:2009}. However, inelastic magnetic response at \textbf{Q}$_1$ does not give evidence of a crossing through a phase transition: no divergence of the static susceptibility $\chi(\mathbf{Q}_1,\omega=0)$ has been reported \cite{Broholm:1991}. Even the shallow maxima of $\chi(\mathbf{Q}_1,\omega=0)$ at $T_0$ may be an artifact of the fitting, or it can also be, as discussed for $\chi(\textbf{Q$_0$},\omega=0)$ a consequence of the Fermi Surface reconstruction.

As recently proposed in two different approaches, DMFT \cite{Haule:2009} and group theory analysis \cite{Harima::2010}, the promising explanation is that the hidden order phase would be a multipolar phase: a hexadecapolar order in DMFT studies, a quadrupolar order in group theory. In this last scenario, the hidden order phase is still hidden as it corresponds to a second order phase transition from the space group $I4/mmm$ (No.139) to the space group $P4_2/mnm$ (No.136) with no lattice distortion and invariance of the Ru-site at the crossover transition from hidden order to antiferromagnetic phase. Switching from the hidden order to the antiferromagnetic phase will preserve the $P4_2/mnm$ (No.136) symmetry of the lattice but will add of course the time-reversal symmetry operator. Let us emphazise that even if the resonance at \textbf{Q}$_0$=(1,0,0) is not a direct proof of the hidden order parameter, there is no doubt that it is a key signature which supports strongly the change from $I4/mmm$ (No.139) to $P4_2/mnm$ (No.136) symmetry at the paramagnetic-hidden order border. If no quadrupolar signature will be detected, a possibility is that the HO phase of \urs\ may be regarded as an electronic spin Peierls transition with only tiny displacement of the atoms ($\delta d/d <10^{-6}$) \cite{fomin}.

\begin{figure}
\begin{center}
\includegraphics[width=72mm]{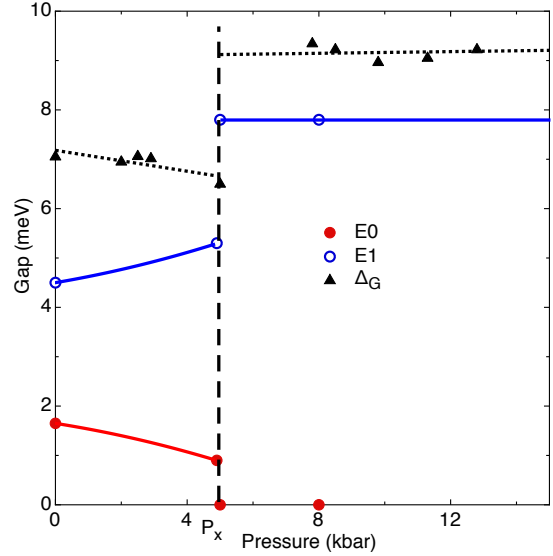}
\end{center}
\caption{Behavior of the resonance energies $E_0$ and $E_1$ versus pressure, above and below $P_x\simeq5$ kbar and of the the gap $\Delta_G$ reflecting the gap opening below $T_0$.}
\label{fig16}
\end{figure}

Furthermore the collection of previous data with fine pressure tuning \cite{Bourdarot:2004b,Bourdarot:2005,villaume:2008} and of recent data at a fixed pressure $P$ between $P_x$ and $P_c$ with supplementary magnetic field $H$ scans\cite{aoki:2009}, allow to extract the $P$ dependence of $E_0$ and $E_1$ and to compare with the $P$ dependence of the gap $\Delta_G$ derived from resistivity measurements (figure \ref{fig16}) \cite{Hassinger:2010,Hassinger:2010b,Jeffries:2007}. This gap $\Delta_G$ is directly related to the partial gap opening at the Fermi surface which occurs at $T_0$.
Using a simple formula $\rho = \rho_0+A\,T^2 + B\,e^{-\Delta_G/T}$ and not the currently used $\rho = \rho_0+A\,T^2 + B\,T/\Delta_G\,(1+2\,T/\Delta_G)\,e^{-\Delta_G/T}$, which is suitable only if the scattering process is due to spin waves, $\Delta_G$ is quite close to $E_1$ by comparison to $E_0$ and comparable to $\Gamma_c$. In the vicinity of $T_0$, as pointed out for example by the entropy drop, the contribution of the $E_1$ resonance may play a major role. However on cooling, the \textbf{Q}$_1$ role is defeated by the \textbf{Q}$_0$ wave-vector response which occurs at lower energy than the $E_1$ one.

Under pressure, as shown in Fig.\ref{fig16}, $E_0(P)$ decreases from 1.8 meV at $P=0$ to 0.8 meV just below  $P_x$ and collapses above  $P_x$ as the response at \textbf{Q}$_0$ is dominated by the onset of a large static sublattice magnetization. On the contrary, for the wave-vector \textbf{Q}$_1$=(1.4,0,0), the resonance at $E_1$ persists through $P_x$; $E_1$ increases smoothly under pressure and exhibits a jump at $P_x$. As for the gap $E_0$, the resonance at \textbf{Q}$_1$ becomes sharp below $T_0$, where the number of carriers drops, but broadens, with a width comparable to $\Gamma_c$ in the over-damped regime. The thermal dependence of the width $\gamma_0$ can be fitted using the Korringa model \cite{Iwasa:2005}: $\gamma_0 \sim \gamma_{0,T=0K} + a\ n(T)^2\ k_BT$ where $n(T)=n_0\,e^{-\Delta(T)/k_BT}$ is the number of carriers which reduced when the gap $\Delta(T)$ opens at the Fermi surface. A gap (temperature independent) $\Delta=7.7$ meV is found below $T_0$ surprisingly very close to our derivation of $\Delta_G$ (70 K) or to the gap value deduced from specific heat measurements ($\Delta_{cp}=73$ K)\cite{vanDijk:1997}.

Thus our data give sound basis for further theoretical developments. They confirm the dual character of the phase transition with $\Delta_G$ directly linked to the itinerant nature of the $5f$ electrons and $E_0$ and $E_1$ collective modes associated to the local character of the $5f$ electrons.
The description of quantities such as $I_{E_0}$ and presumably $I_{E_1}$ as well as the temperature variation of the specific heat below $T_0$ \cite{Balatsky:2009}, by BCS-type formula, may reflect the feedback between the local and itinerant properties. Of course, the possibility that the strong resonance ($E_1$) at \textbf{Q}$_1$ is a mark of an incommensurate spin density wave cannot completely be ruled out.  Our support for the choice of \textbf{Q}$_{AF}$ as the wave-vector of the HO phase are; its occurrence only in the HO phase \cite{villaume:2008}, the quasi-invariance of the frequencies detected in the de Haas-van Alphen effect\cite{Nakashima:2003}  and Shubnikov-de Haas effect through $P_x$ \cite{Hassinger:2010b} that indicate no change of the wave-vector between HO and AF phases, the lower value of $E_0$ by comparison to $E_1$ with furthermore a field convergence of $E_0$ towards $E_1$ in high magnetic field when the intersite dipolar and quadrupolar interaction  are smeared out on entering in the paramagnetic polarized phase where the magnetic response will be $q$ independent \cite{Bourdarot:2003,Levallois:2009}. 

Let us compare the results on \urs\ with 
two Pr skutterudite systems PrRu$_4$P$_{12}$ and PrFe$_4$P$_{12}$ where strong feedbacks occur between Fermi sea and multipole dynamics (see references in \cite{JPSJ77SA:2008}). The interest in the last reference is that the U ions in their tetravalent configuration will have two electrons in the $5f$ shell as do Pr ions in their trivalent configuration in the $4f$ shell.

These are systems where a strong feedback may occur between band structure, charge and multipolar ordering \cite{JPSJ77SA:2008}. In PrRu$_4$P$_{12}$, it is now well established that at low pressure a charge order phase transition at $T_0=63$ K occurs through a switch from $body\ centered\ cubic$ to $simple\ cubic$ lattice with clear evidence of a formation of two sublattice leading here to an unambiguous detection of $Ru-ion$ displacement \cite{Lee:2001,Iwasa:2005}. The strong similarity of PrRu$_4$P$_{12}$ with \urs\ in inelastic neutron scattering experiment is the smearing of the inelastic response above $T_0$ and the appearance of a sharp feature below $T_0$ with nuclear Bragg displacement following BCS-type dependence leading to the claim that the crystal-field level variation through $T_0$ is coupled to the carrier change itself \cite{Iwasa:2005}(as for \urs, the temperature variation of integrated intensity of $E_0$). Another interesting case is PrFe$_4$P$_{12}$ where for $P_x=2$ GPa, the system switches from HO semi-metallic phase to AF insulator phase \cite{Sato:2000}. NMR experiments on P sites \cite{Kikuchi:2007} have recently lead to the conclusion that the HO phase has a scalar OP\cite{Sakai:2007,Kiss:2008}. The difference between \urs\ and the last two skutterudite system appears that for the first case in the paramagnetic regime the system is clearly in a mixed valence state for the U ions (valence v $\sim$ 3.5)\cite{Hassinger:2008a,Barzykin:1995}. As discussed for systems like TmSe \cite{Derr:2006}, it is the crossing to a well ordered phase at $T_0$, which makes that the uranium centers look to be tetravalent and leads consequently to the idea that a ThRu$_2$Si$_2$ description for the electronic bands may be a good starting point \cite{Haule:2009,Harima::2010}.

\section {Conclusion}
The present work leads to a precise study of the resonance $E_0$ at \textbf{Q$_0$} which is up to now the main signature of the OP of the HO phase. The key results are; the control of the temperature dependence of the resonance energy $E_0$ by the partial gap opening at the Fermi surface ($\Delta_G$), the temperature like BCS dependence of the integrated inelastic intensity of the resonance. It was suggested that this variation may reflect the temperature evolution of the order parameter. Clearly, the itinerant and local character of the $5f$ electrons must be treated in equal footing. These new data will certainly push to theoretical developments.

\section{Acknowledgements}

We thank H. Harima, K. Miyake, G. Knebel, L. Malone, M. Zhitomirsky, V Mineev, J.P. Sanchez and J.P. Brison for useful and fruitful comments. 
This work is supported by the Agence Nationale de la Recherche through the ANR contracts Delice, Sinus, and Cormat.

\end{document}